\documentclass{article}

\usepackage{arxiv}

\usepackage[utf8]{inputenc} 
\usepackage[T1]{fontenc}    
\usepackage{hyperref}       
\usepackage{url}            
\usepackage{booktabs}       
\usepackage{amsfonts}       
\usepackage{nicefrac}       
\usepackage{microtype}      
\usepackage{lipsum}

\usepackage{subfig}
\usepackage{graphicx}
\usepackage{algorithm}
\usepackage[noend]{algpseudocode}

\newcommand{\mname}{{\em Conformity}}

\title{Conformity: a Path-Aware Homophily measure for Node-Attributed Networks}

\author{
Giulio Rossetti\\
  KDD-Lab, ISTI (CNR) \\
  G. Moruzzi, 1, Pisa \\
  \texttt{giulio.rossetti@isti.cnr.it} \\
  \And
  Salvatore Citraro\\
  Department of Computer Science, University of Pisa \\
  Largo Bruno Pontecorvo, 3, Pisa \\
  \texttt{salvatore.citraro@phd.unipi.it} \\
  \And
  Letizia Milli\\
  Department of Computer Science, University of Pisa \\
  Largo Bruno Pontecorvo, 3, Pisa \\
  \texttt{milli@di.unipi.it} \\
}

\begin{document}
\maketitle

\begin{abstract}
Unveil the homophilic/heterophilic behaviors that characterize the wiring patterns of complex networks is an important task in social network analysis, often approached studying the assortative mixing of node attributes.
Recent works underlined that a global measure to quantify node homophily necessarily provides a partial, often deceiving, picture of the reality.
Moving from such literature, in this work, we propose a novel measure, namely \mname, designed to overcome such limitation by providing a node-centric quantification of assortative mixing patterns.
Differently from the measures proposed so far, \mname\ is designed to be path-aware, thus allowing for a more detailed evaluation of the impact that nodes at different degrees of separations have on the homophilic embeddedness of a target.
Experimental analysis on synthetic and real data allowed us to observe that \mname\ can unveil valuable insights from node-attributed graphs.
\end{abstract}


\footnotetext{2020 IEEE. Personal use of this material is permitted. Permission
from IEEE must be obtained for all other uses, in any current or future
media, including reprinting/republishing this material for advertising or
promotional purposes, creating new collective works, for resale or
redistribution to servers or lists, or reuse of any copyrighted
component of this work in other works.}

\section{Introduction}
\label{sec:intro}
During the last decades, network science has become one of the fastest growing multidisciplinary research fields.
Every year, countless researchers, from heterogeneous backgrounds, leverage network theory to analyze complex data describing alternative facets of real world phenomena.
From sociology to biology, more and more domains study entities composed of several components - each having its internal complexity and peculiar functionalities - all of them strictly tied in functional relationships.
Such complex organizations can naturally be modeled as networks, and as such, analyzed. 
While reasoning on networks built on top of contextual data, topology is only one of the aspects to take into account: nodes and edges often carry additional semantic information that are of uttermost importance to properly understand the phenomena expressed by the underlying topological structure.
Often, such augmented structures are referred to as \textit{Feature-rich networks} \cite{Interdonato2019}.
That general term acts as an umbrella for several, more specific, class of network extensions including temporal as well as probabilistic and attributed (or labeled) networks.
In this work, we are particularly interested in \textit{labeled} or \textit{node-attributed networks}, where reliable external information is added to the nodes as categorical or numerical attributes.
Node-attributed graphs are a quite expressive model of social network environments since several salient dimensions (age, gender, nationality\dots) can be meaningfully studied by leveraging such a framework.

Indeed, one of the salient aspects that makes network science a widespread research methodology is its ability to unveil emergent behaviors of complex systems.
Network topology is, perhaps, the clearest example of how the overall complexity of a whole system is more than the sum of the coupled interactions among its components. 
Several modeling works have shown how some universal network properties are the results of emergent behaviors: classic examples are the long-tail degree distribution \cite{barabasi1999emergence} and the meso-scale modular organization \cite{fortunato2016community} that describe complex systems as sparsely connected dense components.
Another relevant emerging behavior is \emph{homophily}. 
It has been observed that individuals are more likely to group in social circles if they share common features and stay apart when some specificity diverges.
Social network analysis has deeply investigated such a phenomenon, trying to measure its impact and propose a mechanistic explanation to its existence.
A proxy often used to estimate for homophilic behaviors fall under the name of Newman's \emph{assortativity} \cite{newman}.
Such a measure aims to classify a whole network in a range that goes among two extremes: \emph{disassortative mixing}, where nodes are likely to be connected if they are anti-correlated w.r.t. a given property, and \emph{assortative mixing}, where, conversely, nodes are likely to be connected if they share a given property.
Assortativity has been widely studied and applied to characterize several phenomena such as degree correlation and node-attribute correlations.
One of the major drawbacks of such a measure, and similar ones, lies in its definition scale: a complex behavior is summarized in a single, average, score.
Recently, a few works \cite{peel} tried to overcome such limitation by proposing a multiscale extension of Newman's assortativity, thus allowing to analyze multimodal behaviors that the original score makes impossible to observe (e.g., identifying different, even conflicting, homophilic/heterophilic behaviors within the same complex system).

In this work, we move from such a line of research, proposing an alternative proxy for measuring multiscale node homophilic couplings: \mname, a node-centric path-aware measure, able to unveil heterogeneous mixing patterns in node-attributed networks, designed to cope with categorical (single and multi)-attributes.
Inspired by a higher-order assortativity definition, namely the \textit{clumpiness} score \cite{estrada2008clumpiness}, \mname\ takes into consideration the evidence that nodes with similar characteristics are not divided by long chains.
Experimental results carried out on real world node-attributed networks underline that \mname\ allows to study homophilic patterns from a novel point of view and make valuable inference on the social contexts it is applied to.



The work is organized as follows.
Section \ref{sec:related} introduces the relevant literature to frame the proposed contribution; Section \ref{sec:def} formally introduces \mname; Section \ref{sec:analysis} discusses experimental results obtained applying \mname\ to synthetic as well as real-world data. 
Finally, Section \ref{sec:disc} concludes the paper.

\section{Related}
\label{sec:related}

Literature defines social homophily as the tendency of people to interact with similar others in respect of dimensions such as age, gender, education, as well as values, attitudes, and political beliefs, sourced by geographical distances, households, workplaces, and universal human cognitive processes \cite{mcpherson2001birds}.
Hidden social dynamics can be unveiled studying homophily as well as heterophily among people: in the presence of segregation, interracial friendships are less probable when social class is correlated with race \cite{moody2001race}; in the early school grades, boys tend to form larger and more heterogeneous cliques compared to the smaller and more homogeneous cliques of girls \cite{shrum1988friendship}; intergroup mixing is also a key factor in academic success when interdisciplinary research is involved \cite{feng2020mixing}. 
Such a brief set of examples let us know how both homophily, and its counterpart, act as fundamental principles in the choice of people's social circles.

In the language of network science, they act as a discriminant factor for node neighborhood selection.
Network homophily can refer either to explicit topology (e.g., nodes with a similar degree preferably connect) or to the interactions between nodes sharing similar labels.
\textit{Newman's assortativity coefficient} \cite{newman} is the most known and used measure for quantifying homophily in complex networks.
Based on modularity, the coefficient is calculated as the sum of the differences between the observed and the expected fraction of edges between nodes sharing similar values of an attribute.
Some recent extensions or alternative approaches, like \textit{ProNe} \cite{rabbany2017beyond} or the \textit{VA-Index} \cite{pelechrinis2016va}, are also able to cope with pairs of attributes or vector of features, shedding light, more than \textit{Newman's coefficient}, on the phenomenon of similarity between two or more attributes based on network structure.

Such global and aggregated measures flatten and simplify a heterogeneous context in one only score, and avoid the presence of outliers or different mixing interactions characterizing different zones of networks and perhaps also single nodes.
In such scenarios, local or node-centric approaches (able to assign a score to each graph node) should help for quantifying a more reliable and exploitable network description.
Since the only direct neighborhood (or ego-network) of nodes can not be taken into consideration due to its limited expressive power (inherited in the scale-free-like degree distribution of complex networks), the issue is to define connectivity boundednesses able to circumscribe those nodes whose importance is fundamental in the assortative attitude measurement of a target one.
While some lines of research focused on degree assortativity \cite{noldus2015assortativity} (extended to cope with higher-order notions of node neighborhood such as a two-walks degree correlation \cite{allen2017two} or transsortativity \cite{ngo2020transsortative}), the node-attributed counterpart of the problem has not received much of attention.

Only a few studies address such a task in this latter scenario.
Recent works aimed to study the existence of possible relations among network structure and label distribution among nodes (e.g., how structure and minority size generate perception biases \cite{lee2017homophily}) as well as shed light on the individual differences in mixing (e.g., in the analysis of \textit{monophily}, a concept aiming to identify those individuals with extreme preferences for different labels \cite{altenburger2018monophily}).
Accordingly, inferring and quantifying individual differences as well as different local mixing comes as a hard task in complex networks studies.
A model able to characterize the within-group mean and variation of mixing patterns was recently proposed in the framework of Bayesian inference \cite{cantwell2019mixing}:
when variation is consistently present, the group mean only is not able to fully describe individual node preferences.
In some work, \textit{locality} is exploited through a definition of assortativity based on the correlation between two consecutive nodes visited by a random walker.
For instance, this rationale is used in \cite{gutierrez2019multi}, and applied in the graph classification task;
a multi-hop assortativity is defined, here, as the probability that a randomly selected node and a randomly selected $t$-hop neighbor belong to the same category, where $t$ indicates the time of the visit of the random walker.
Closer to the current work, a node-centric and Newman's-normalized measure, namely \textit{Peel's assortativity} \cite{peel}, was recently proposed in the context of local-aware homophily, modeling similarities between nodes as an autocorrelation of a time-series defined as a sequence of node labels visited by a random walker with restart.

\section{\mname}
\label{sec:def}

We aim to design a local proxy to measure the degree of homophilic embeddedness of network nodes w.r.t. the attributes they carry.
Such a task has been recently approached by Peel et al. \cite{peel} to overcome the limitation of classical approaches that usually propose a single aggregate score to characterize the overall assortativity of network nodes.
A multiscale strategy to estimate the presence of homophilic patterns within a complex system enables the discovery of emergent behaviors that classical indexes often are not capable of unveiling.
The score proposed in \cite{peel} moves from the classical Newman's assortativity \cite{newman} that, in turn, poses its ground on a reinterpretation of the modularity score - a measure often used to quantify the quality of network clustering partitions.

Modularity, $Q$, computes the difference between the observed and the expected fraction of edges between nodes sharing similar attribute values: in the assortativity coefficient, $r_{global}$, such quantity is normalized in the range $-1\leq r_{global} \leq 1$. 
Thus: $r_{global}=1$ implies that all edges only connect nodes labeled with the same value; $r_{global}=0$ that all edges are randomly connected, and; hypothetically, $r_{global}=-1$ that all edges only connect nodes with a different value.
Formally,
\begin{center}
$r_{global} = \frac{Q}{Q_{max}} = \frac{\sum_{g}{e_{gg}} - \sum_{g}{a^{2}_g}}{1- \sum_{g}{a^{2}_g}}$
\end{center}
where $e_{gg}$ is the proportion of edges connecting nodes of the same type $g$, and $a_{g}=\sum_{i\in g}{k_i/2m}$ is the sum of degrees ($k_i$) of nodes with type $g$.

Indeed, the approach in \cite{peel} yields valuable results; however, it misses a fundamental high-order property of networks: the length of paths connecting nodes. 
To address such an issue, we define a novel measure, namely \mname \footnote{Python code available at \\ https://github.com/GiulioRossetti/conformity}.
\smallskip

Given an undirected attributed network $G=(V,E,A)$, where $V=\{v_1, v_2, \dots, v_n\}$ is the set of nodes, $E=\{(v_i,v_j)|v_i,v_j \in V\}$ the set of edges among them, and $A=\{l_1,l_2,\dots, l_n\}$ the set of node attributes, \mname\ computes the similarity between the attributes of the node $u\in V$ with the ones of the other nodes of the network, weighing it with the distance among them. 
Here, we will focus only on networks with nodes carrying categorical attributes.

\begin{figure*}[t!]
\centering
 \subfloat[]   {\includegraphics[scale=0.13]{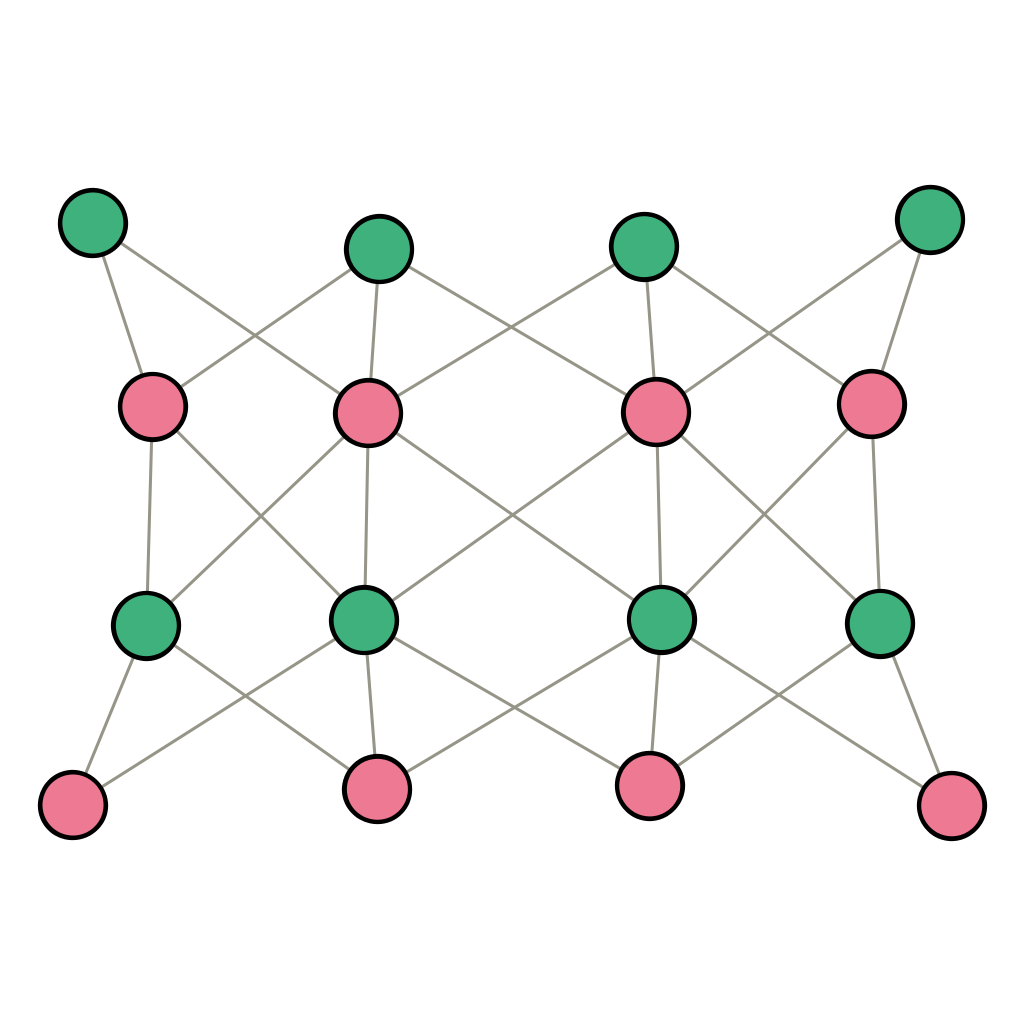}}
  \subfloat[]  {\includegraphics[scale=0.13]{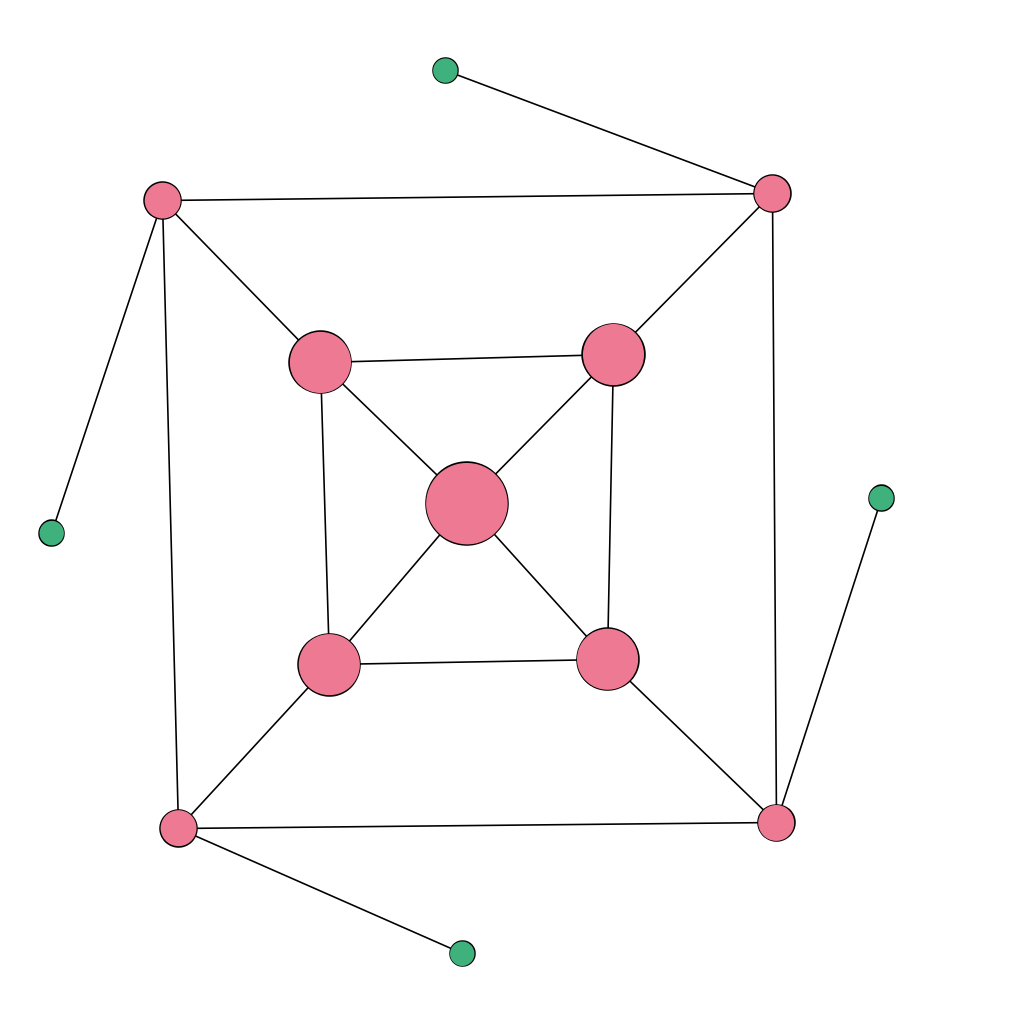}}
    \qquad \qquad
   \subfloat[] {\includegraphics[scale=0.35]{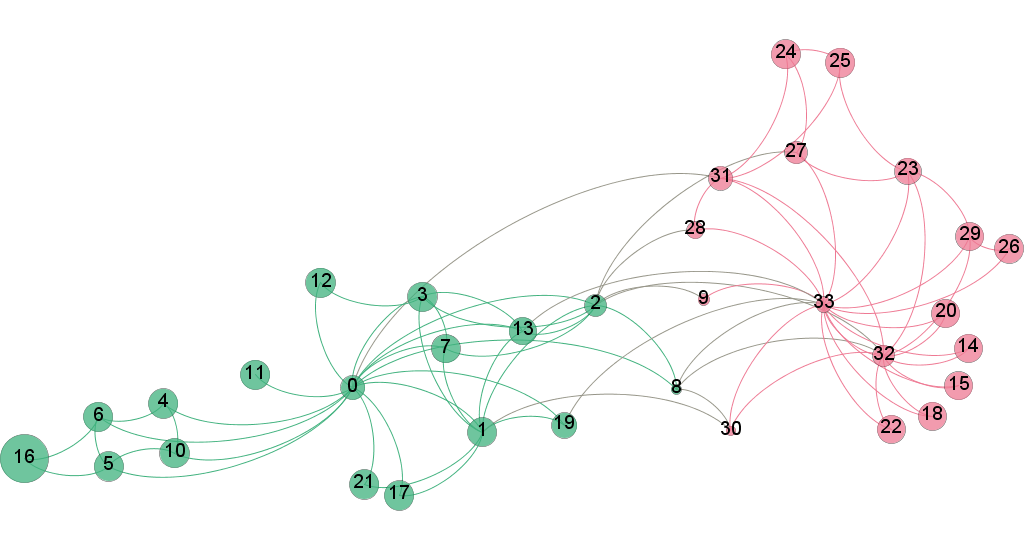}}
    \caption{{\em Toy Examples.} (a-c) Node colors map categorical attribute values, while node sizes encode the respective \mname\ scores (the smaller the size, the lower the score). (a) A scenario in which all nodes minimize the \mname\ score: all nodes have the same size, $\psi(u,\alpha)=-1$, since no connection exists among pairs sharing the same color. (b) The effect of distance on the $\psi(u,\alpha)$ value: the central node's score approaches 1, while moving toward the graph periphery (composed by nodes having different color) nodes' $ \psi(u,\alpha)$ decreases -- reaching negative values for the extreme periphery. (c) {\em Karate Club.} Node colors encode the two factions of the Karate Club dispute, node sizes are proportional to \mname\ scores for $\alpha=2.5$ }
    \label{fig:toy_limit}
\end{figure*}

To facilitate the introduction of \mname\, we need to define a few support functions.

Considering a node $u\in V$ we define the set $N_{u,d}$ as the set of $u$ neighboring nodes at a distance $d$:
\begin{equation}
N_{u,d} = \{v| dist(u,v)=d\}.
\end{equation}
Moreover, lets call $I(u,v)$ the indicator function that compares the attribute values of two nodes $u,v \in V$
\begin{equation}
    I_{u,v} = 
\left\{
    \begin{array}{ll}
        1  & \mbox{if } l_u=l_v \\
        -1 & \mbox{otherwise}
    \end{array}
\right.
\label{eq:Iuv}
\end{equation}

and $f_{u,l_u}$ the function that, if among the neighboring nodes of $u$ there is at least one node sharing the same attribute value $l_u$, computes the ratio of $u$'s neighbors sharing it
\begin{equation}
    f_{u,l_u} = \frac{|\{v|v\in \Gamma(u) \land l_u=l_v\}|}{|\Gamma(u)|},
\end{equation}
where $\Gamma(u)$ is the first order neighborhood of node $u$, i.e., the set of nodes adjacent to it.
Moreover, to assure a consistent interpretation of \mname, we force $f_{u,l_u}$ to assume values in $(0, 1]$ by setting its value to 1 when its numerator nullifies.

\noindent Finally, we define the \mname\ score for a node $u\in V$ and a given real number $\alpha$ in $[0, +\infty)$ as:

\begin{equation}
    \psi(u,\alpha) = \frac{\sum_{d\in D} \frac{\sum_{v \in N_{u,d}} I_{u,v} f_{v,l_v}}{|N_{u,d}| d^\alpha}}{\sum_{d\in D} d^{-\alpha}},
    \label{eq:clumpiness}
\end{equation}
where $D$ is $max(\{dist(i,j)|i,j \in V\})$, and the parameter $\alpha$ controls the level of interaction between nodes, which exponentially decreases while the distances among nodes increase;
thus, imposing $\alpha=1$, we force a linear decrease w.r.t. the distance, while $\alpha>1$ imposes a sublinear decrease which reduces the level of interaction between relatively distant nodes.

\mname\ can be algorithmically interpreted as follows. 
\begin{enumerate}
    \item For each node pair $u,v\in V$, with $v\in N_{u,d}$  with $1\le d \le max(\{dist(i,j)| i,j\in V\})$ the nodes attribute concordance -- given by $I(u,v)$ -- is weighted by $f_{v,l_v}$, namely the degree of homophily of the node $v$ toward its first order neighborhood;
    \item The average of such score aggregated over all the nodes in $N_{u,d}$ is then damped by a factor $d^\alpha$, to account for the distance that separates the nodes considered by the source $u$.
    Note that we used an inverse power-law distance decay -- that recalls well-known physical measures such as the Coulombic and gravitational ones -- since such an approach has already proven its consistency in the definition of the clumpiness measure \cite{estrada2008clumpiness}, a widely used degree dispersion index.
    \item Finally, the computed score is normalized to ensure that \mname\ lies in the range $[-1, 1]$.
\end{enumerate}

\smallskip

Intuitively, the value of $\psi(u, \alpha)$ is maximized when a node $u$ is surrounded by neighbors having the same attribute value, minimized in the opposite scenario.
Fig. \ref{fig:toy_limit}(a) shows a network whose nodes (colored by their attribute value) always minimize their \mname\ value independently from the chosen decay exponent. 
Such a limit case example perfectly captures the essence of anti-conformity: edges always connect nodes with a different attribute value, resulting in the absence of homophilic islands.
Conversely, Fig. \ref{fig:toy_limit}(b) shows a simple scenario where the length of the paths among nodes sharing different labels plays a crucial role in the \mname\ values.
We can easily observe how \mname\ (coded with the relative node size) tends to decrease moving from the inner layer to the outer ones -- e.g., moving from the more homophilic embedded nodes to the more heterophilic ones.
\smallskip

As discussed, \mname\ is a node-related measure: we can define the overall degree of \mname\ of a network as:
\begin{equation}
    \Psi(\alpha) = \frac{1}{|V|}\sum_{u\in V} \psi(u, \alpha).
\end{equation}
Indeed, such average score is only able to capture a general trend, not to provide a clear picture of the emergent homophilic behaviors at a local level.
\smallskip

To better understand the information that the proposed measure can unveil, let us consider the classic example offered by Karate Club dataset \cite{zachary1977information}, representing the small social network of a club after a conflict arose between the administrator, ``John A.", and an instructor, `Mr. Hi".
The graph is classically used as a toy example for characterizing community discovery algorithms since it is neatly divided into two factions and very suitable for explaining a clustering methodology.
Moreover, since each node is labeled with the club it belongs (``John A." or ``Mr. Hi"), this external information is commonly exploited as a ground truth to test the goodness of the algorithm outputs, even if it has been shown not to be a proper approach \cite{peel2017ground}.
In Fig. \ref{fig:toy_limit} (c), different colors encode the two categorical node attribute values characterizing the network while, as in the previous example, the node sizes are proportional to the node \mname\ score ($\alpha=2.5$). 
As we expected, the highest \mname\ values are assigned to those nodes that prevalently connect to same attributed peers while, on the other hand, the lowest ones characterize bridge-nodes.  
Particular attention must be paid to node 8, which registers the lowest \mname\ score ($\simeq$-0.18).
Indeed, the data paper that discusses the origin of the Karate Club network dataset \cite{zachary1977information} help us in providing a neat justification for such \mname\ value: node 8 identifies a weak supporter of ``Mr. Hi", that joined with the ``John A."'s faction, after the split, for personal advantage, so he represents a bridge between the two opposite sides of the Karate Club dispute indeed. 

\begin{figure*}[t!]
\centering
    {\includegraphics[scale=0.22]{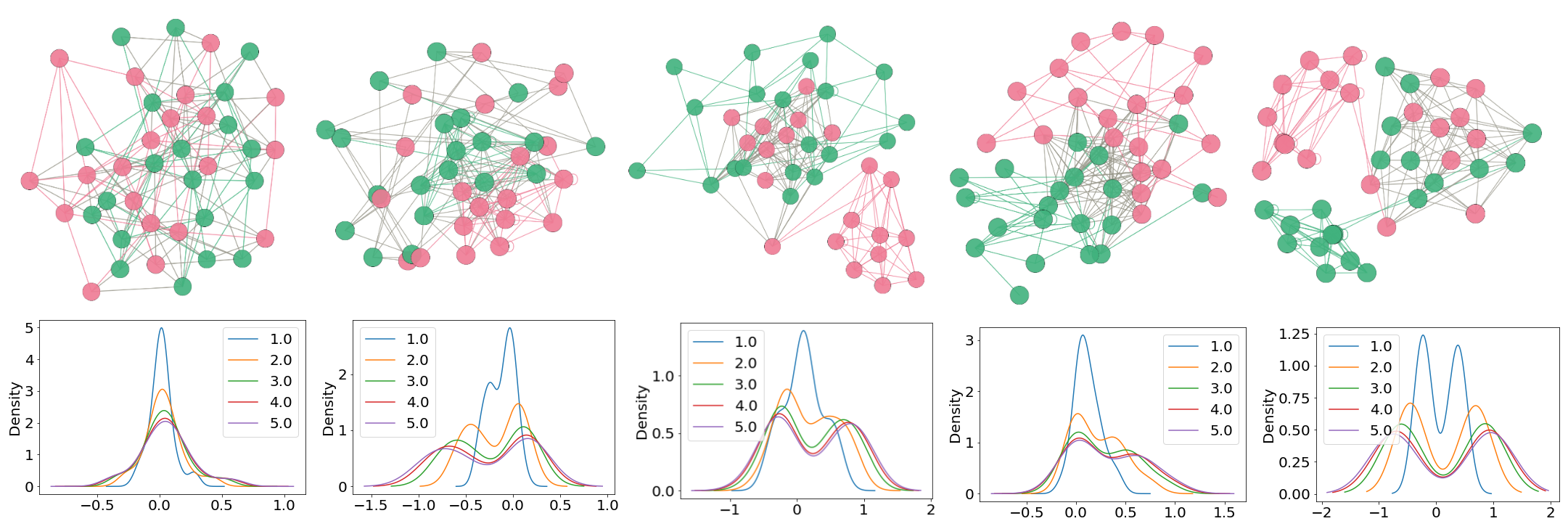}}
    \caption{{\em Peel's quintet toy example.} KDE's distributions of several local mixing patterns according to \mname, for different values of $\alpha$: the higher the value, the less the contribution of distant nodes to the target final score -- as shown by the progressive amplification of the distributions toward close-to-bound values.}
    \label{fig:toy_net}
\end{figure*}

\section{Experimental analysis}
\label{sec:analysis}

Studying the homophilic patterns of actors embedded in a network is a way to unveil emergent behaviors that are otherwise hard to identify.
In this section, we propose a characterization of both synthetic and real-world networks using the proposed \mname\ score.

\smallskip 

\begin{figure*}[t!]
\centering
    \subfloat[] {\includegraphics[scale=0.34]{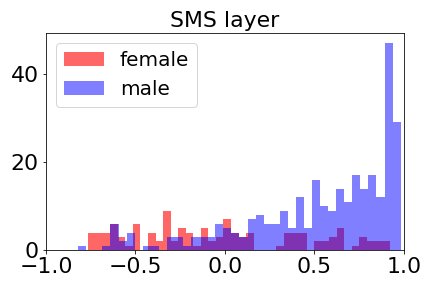}}
    \subfloat[] {\includegraphics[scale=0.34]{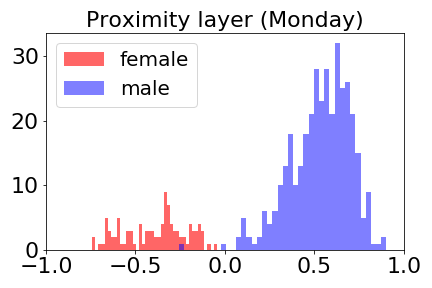}}
    \subfloat[] {\includegraphics[scale=0.34]{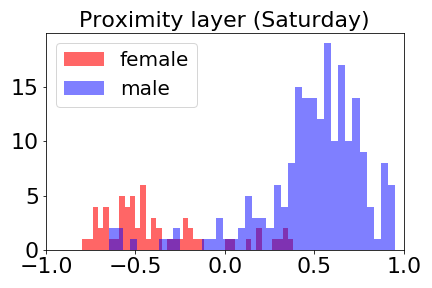}}
    \caption{\textit{Copenhagen Network analysis.} (a) \mname\ analysis ($\alpha=2.5$) in the SMS layer analysis; (b-c) \mname\ analysis ($\alpha=2.5$) in the proximity graphs of Monday and Saturday.}
    \label{fig:cop}
\end{figure*}
\subsection{Synthetic data} 
Inspired by the Peel's quintet \cite{peel}, in Fig. \ref{fig:toy_net} we replicate the building of a set of five small synthetic graphs with the same number of nodes and edges (40 nodes -- 20 red, 20 green -- and 160 edges), but involving a rewiring of edges leading to the emergence of different local mixing patterns that Newman's assortativity coefficient, $r$, is not able to detect (i.e., $r=0$).
Indeed, Newman's score is a valid indicator only for the leftmost graph of the figure, the only one where all edges are randomly rewired across all nodes.
This is showed by the unimodal distribution in the \mname\ plot for (a). In contrast, other plots reflect and capture the heterogeneous patterns obtained by planting homophilic relations among nodes: in such scenarios, the unimodal distribution breaks down into bimodal ones -- e.g., the twin peaks observed for the rightmost graph describe the most extreme case where exactly half the nodes is perfectly homogeneous; in contrast, the other half is entirely heterogeneous.

An aspect worth noticing is the effect played by the $\alpha$ parameter on the $\psi$ values.
As discussed, the $\alpha$ exponent allows tuning \mname\ sensitivity w.r.t. the distance among node pairs. 
For $\alpha=0$, all nodes are perceived at the same distance from the source node, thus contributing equally to its final score; for $\alpha>0$, the contribution of nodes is weighted w.r.t. their distance, and progressively dumped while increasing such value.
The effect of increasing $\alpha$, as shown by the KDEs distributions in Fig. \ref{fig:toy_net}, is to concentrate the actual contribution to low-distance neighborhoods, thus favoring a polarization of the scores to the extreme values of the domain.
Indeed, there is no one-fits-all value for such parameter: it needs to be fitted to the analytical needs and the underlying network topology.


\subsection{Real data} 
\noindent {\bf Copenhagen Network Study.} 
We firstly consider a small real-world network, namely the interaction data from Copenhagen Network Study \cite{sapiezynski2019interaction}.
It is composed of different layers connecting a sample of 700 among male and female university students for four weeks: we consider, here, the SMS layer and the proximity estimated via Bluetooth signal strength.
Since information about node gender is available, we mainly aim to relate a characterization of the network based on \mname\ to some of the analysis already shown in the original data paper, e.g., more frequent male-male interaction than male-female and female-female ones \cite{sapiezynski2019interaction}. 
Since the underlying network reflects these frequencies, we describe homophily by gender leveraging \mname, trying to give more insights than the only number of exchanged messages.
Fig. \ref{fig:cop} shows that several male nodes are perfectly homophilic w.r.t. gender, but also that there exist a few highly heterophilic ones among them.
The same (i.e., the same mixing pattern) is not true observing female node \mname\ distribution, even taking into account the fact that the two populations are unbalanced.
Considering the proximity layer, we show the graph analysis of two days, namely Monday and Saturday.
Fig. \ref{fig:cop}(b-c) underlines how different mixing patterns arise considering different days of the week.

\smallskip

\noindent {\bf Facebook100.} 
Facebook100 \cite{Traud100Facebook} is a collection of 100 Facebook friendships networks among 100 U.S. colleges, built during the early history of the social network.
Nodes are labeled with several categorical attributes, profiling people by gender, college year, dormitory\dots
In the following, we will focus on the first 50 networks ordered by size, considering two single-attributes analyses - namely gender and college year - and a multi-attribute overview.
Be aware that the gender attribute yields three values, referring to male, female and missing information; quoting the original data paper, \textit{we use a ``missing" label for situations in which individuals did not volunteer a particular characteristic} \cite{Traud100Facebook}, namely that the individual itself does not specify his gender.

\smallskip

\begin{figure*}[t!]
\centering
    {\includegraphics[scale=0.36]{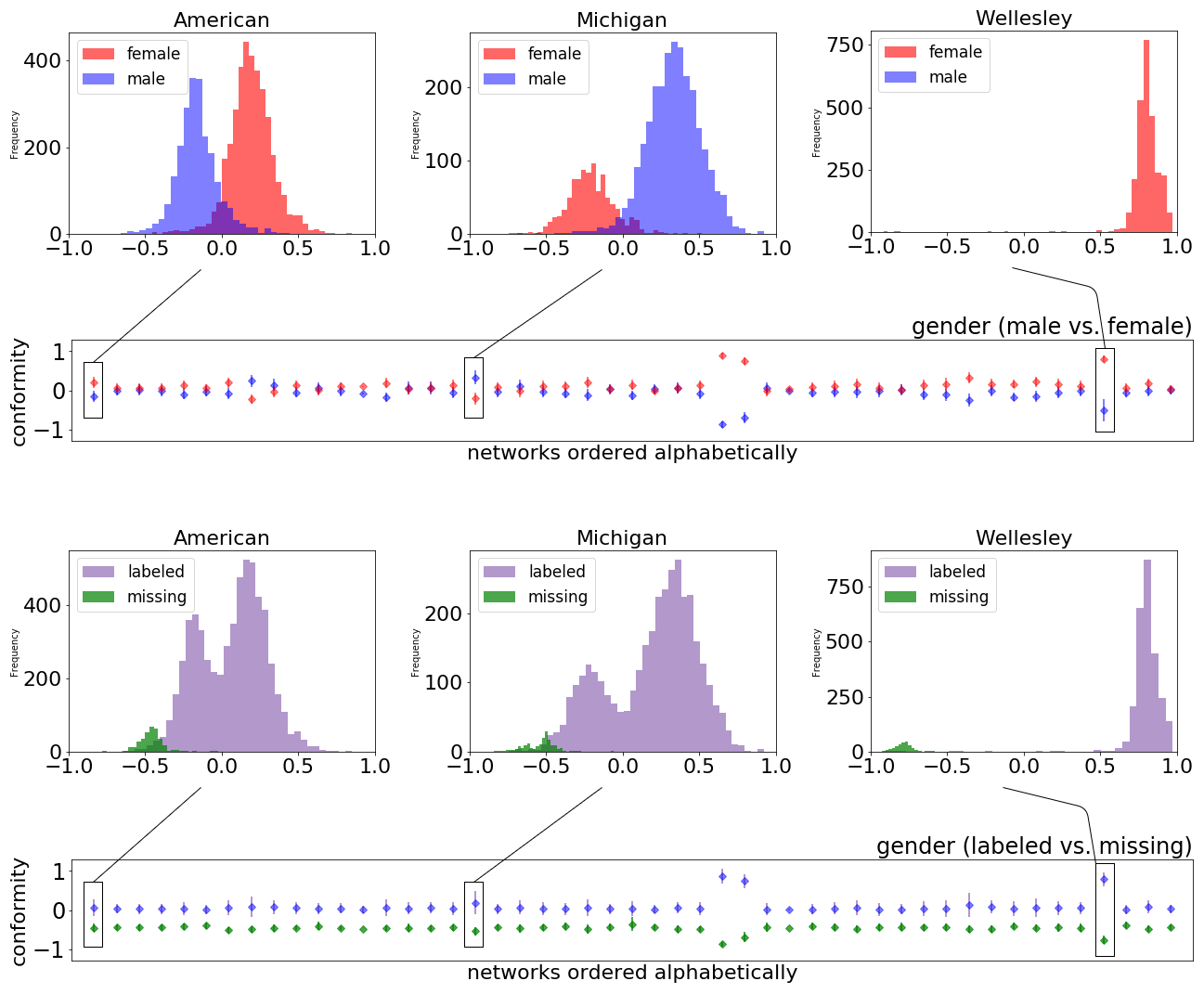}}
    \caption{{\em Gender analysis} ($\alpha=2.5$).
    The box-plot above compares male (blue diamonds) and female (red diamonds) distributions of the analyzed colleges, while the box-plot below compares male-and-female (purple diamonds) and missing values (green diamonds) distributions. Three binned networks show heterogeneity of distributions along the colleges.}
    \label{fig:gender}
\end{figure*}

\begin{figure*}[t!]
\centering
    {\includegraphics[scale=0.34]{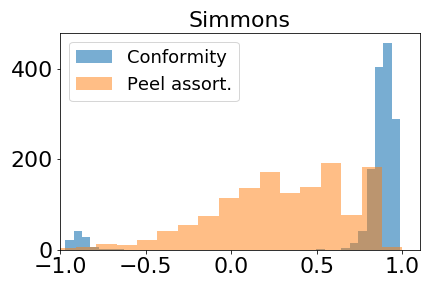}}
    {\includegraphics[scale=0.34]{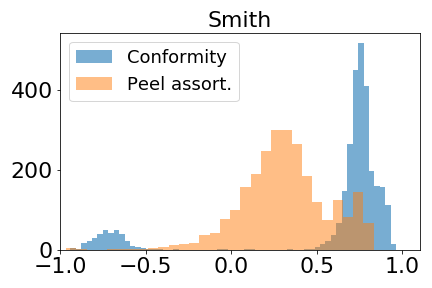}}
    {\includegraphics[scale=0.34]{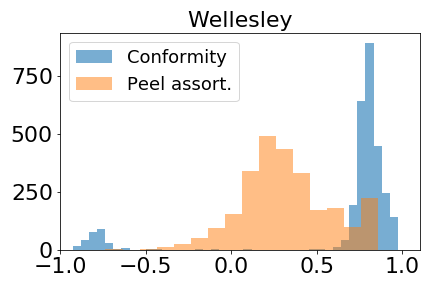}}
    \caption{A comparison between \mname\ ($\alpha=2.5$) and \textit{Peel's assortativity} in the three female colleges.}
    \label{fig:gender_compare}
\end{figure*}

\noindent{\em Gender.} Fig. \ref{fig:gender} shows gender assortativity of the 50 selected networks.
In general, we can not state a male/female tendency to homophily/heterophily as a common behavior across all networks; even if it seems that females' average behavior is more assortative than males, this should be examined on a case-by-case basis.
Nevertheless, for the work, it is more interesting to focus on the scoring of three specific networks, whose male and female homophilic behaviors are different w.r.t. the other colleges.

They are \textit{Simmons}, \textit{Smith} and \textit{Wellesley}, whose distributions are also highlighted in Fig. \ref{fig:gender_compare}, in view of a comparison with \textit{Peel's assortativity} \cite{peel} \footnote{This other measure has also a parameter $\alpha \in [0,1]$, which is integrated over all its possible values in the paper where it is defined and presented \cite{peel}: we replicated the same approach for the comparison in the current study.}.
First of all, referring to the analysis present in the original data paper \cite{Traud100Facebook}, they are three predominantly female colleges whose \textit{Newman's assortativity coefficient} tends to 0.
Leveraging \mname, we can observe (Fig. \ref{fig:gender}) how i) the few male nodes connect disassortatively by gender (i.e., form ties only with females), inducing the emergence of two extreme and distinct mixing, and meanwhile ii) we observe some differences with \textit{Peel's assortativity}, where the same overall strong assortative behavior of the networks is not maintained (Fig. \ref{fig:gender_compare}).
Apparently, the extreme disassortative behavior of few nodes should not so strongly affect the entire network mixing.
Since a real comparison between the two measures is not possible i) due to the absence of ground truth, but mostly because ii) they capture different aspects of mixing, our interpretation is that the local assortativity variant we face suffers from the same limits about network constraints impacting on the reaching of the whole measure range, as already studied in \cite{cinelli2019network}.

Also, the presence of missing values has a non trivial effect on the resulting \mname\ distribution.
The ability to discriminate noisy information from sensible one is important while analyzing a complex system. Since nodes with missing information are homogeneously distributed within the network tissue, \mname\ can correctly classify them as noise, as shown in the \textit{labeled vs. missing} box-plot of Fig. \ref{fig:gender}.
This observation simply implies that these nodes can not induce to homophilic behaviors since missing information is not a real social dimension implying assortative attitudes.

\begin{figure*}[t!]
\centering
    {\includegraphics[scale=0.33]{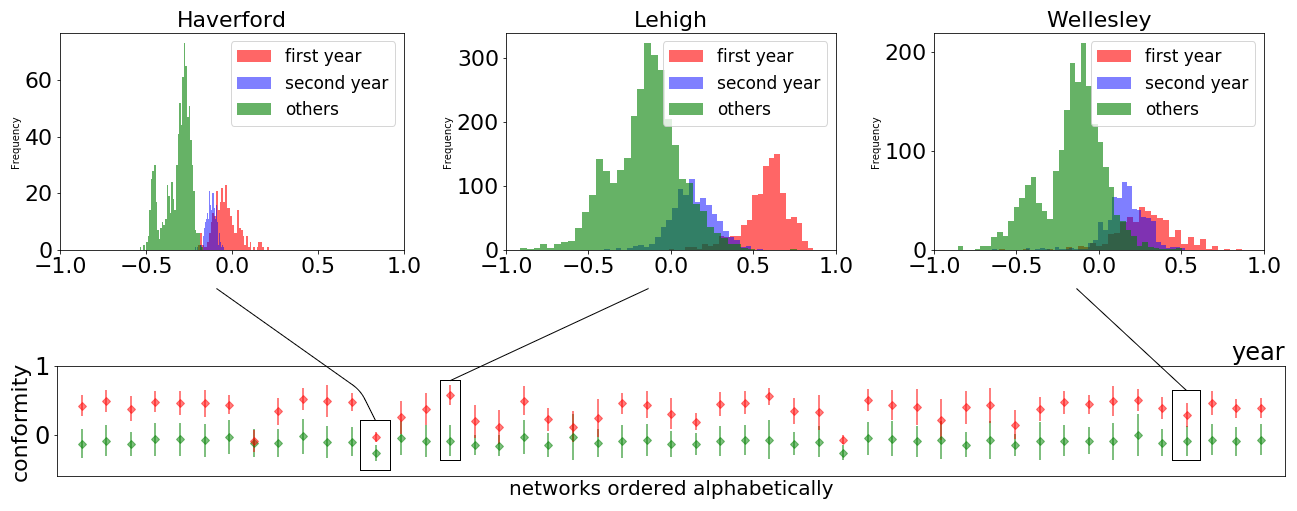}}
    \caption{{\em Year analysis} ($\alpha=2.5$).
    The box-plot compares the first year (red diamonds) and other years (green diamonds) distributions of the selected colleges. Three networks are selected, where also a distinction between first and second year students is highlighted.}
    \label{fig:year}
\end{figure*}

\smallskip

\noindent{\em Year.}
Fig. \ref{fig:year} shows year assortativity of the 50 selected networks.
As already shown in \cite{peel}, first year students highly contribute to the homophilic behavior of the attribute, even when the network attitude does not tend to be globally assortative (see \textit{Haverford} in Fig. \ref{fig:year}).
According to the original data paper \cite{Traud100Facebook}, the year attribute is the most assortative in terms of Newman's coefficient.
Also, in this case, a node-centric measure tends to discover different mixing pattern and allows to differentiate the values that show high homogeneity from the ones that prefer a heterogeneous neighborhood; the three binned networks in Fig. \ref{fig:year} also suggest that homophilic behavior tends to decrease with the increase of enrolment years in a coherent way with the ordinal nature of the attribute.
\smallskip

\begin{figure*}[t!]
\centering
    {\includegraphics[scale=0.33]{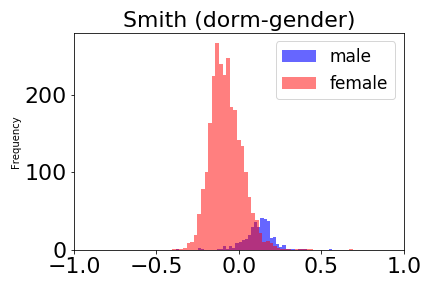}}
    {\includegraphics[scale=0.33]{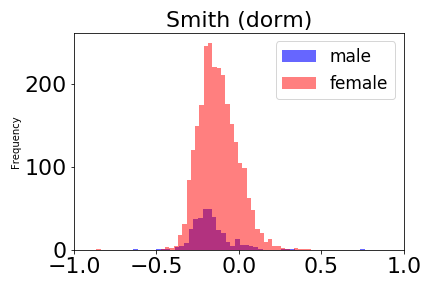}}
    {\includegraphics[scale=0.33]{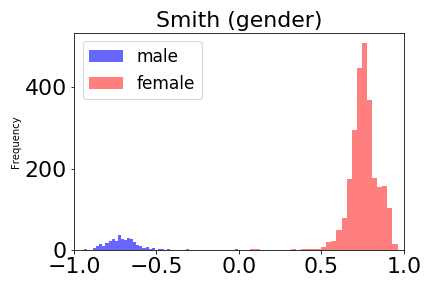}}
    {\includegraphics[scale=0.33]{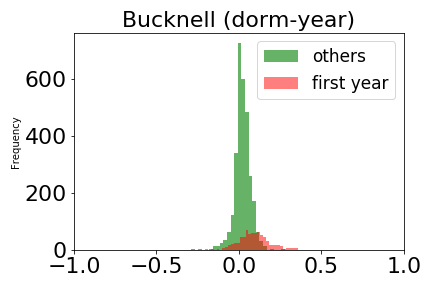}
    {\includegraphics[scale=0.33]{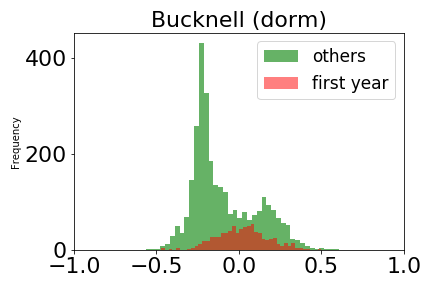}}
    {\includegraphics[scale=0.33]{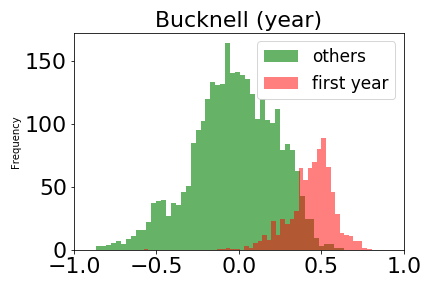}}
    \caption{{\em Multi-attribute} ($\alpha=2.5$). Dorm-gender and dorm-year analysis of \textit{Smith} and \textit{Bucknell} colleges: respectively, \textit{male-female} and \textit{first year vs. other years} differences are highlighted in the distributions.}
    \label{fig:multiattr}}
\end{figure*}

\noindent{\em Multi-attribute.} In a multi-attribute scenario, we want to measure homophily among complex node profile composed by multiple independent fields.
Fig. \ref{fig:multiattr} focuses on dorm-gender and dorm-year assortativity of two selected networks.
\textit{Smith} college, as \textit{Wellesley}, shows a consistent difference between male-female distributions when the only gender attribute is considered, while no substantial differences are highlighted when the only dormitory attribute is analyzed.
However, male nodes tend to be more assortative than female ones when the two attributes are measured together, allowing us to provide a more reliable description of the social media friendships mirroring college interactions. 
Like all other colleges, first year students are highly assortative w.r.t. the other years, while the same pattern does not emerge considering the dormitory attribute.
However, such a pattern emerges anew when dormitory and years are analyzed in a multi-attribute scenario.

\section{Discussion and future work}
\label{sec:disc}

This work introduced \mname, a novel strategy to measure the homophilic mixing of network nodes w.r.t. their categorical attributes.
The proposed measure aims to address some limitations of the well-known assortativity coefficient, in its classic definition given by Newman's work \cite{newman}.
The main reason behind \mname\ is the need to take into account (the often neglected) impact of node distance on the homophilic/heterophilic behaviors that, in social contexts, favor the creation of social ties.
As shown, the proposed measure can unveil interesting nodes' behaviors and can, in practice, be fruitfully adapted to support several tasks (e.g., the identification/measuring of echo-chambers or polarized islands among users living in a social media ecosystem).

In particular, the multi-attribute analysis it enables can support fine grained analysis of complex homophilic patterns to uncover, for instance, homogeneous nuclei among individuals w.r.t. their age and political views, thus supporting tasks such as attributed community discovery \cite{citraro2020identifying}.
Moreover, \mname\ ability to characterize different extreme behavior of even handfuls of nodes (as seen both in homophily by gender analysis of colleges as \textit{Smith} and \textit{Wellesley} and in noise isolation when in the presence of missing values) is a promising feature that can support a wide set of network related task as, for instance, graph-based anomaly detection. 

As future works, since in the current study we focused only on networks encoding categorical attributes, we plan to extend \mname\ to handle scalar attributes.
We also plan to propose an approximate version of \mname\ to lower its computational complexity and to study its effectiveness as support for network analysis tasks in heterogeneous applicative scenarios.

\subsection*{Acknowledgments}
This work is supported by the scheme 'INFRAIA-01-2018-2019: Research and Innovation action', Grant Agreement n. 871042 'SoBigData++: European Integrated Infrastructure for Social Mining and Big Data Analytics'.


\bibliographystyle{ieeetr}
\bibliography{references.bib}

\begin{thebibliography}{10}

\bibitem{Interdonato2019}
R.~Interdonato, M.~Atzmueller, S.~Gaito, R.~Kanawati, C.~Largeron, and A.~Sala,
  ``Feature-rich networks: going beyond complex network topologies,'' {\em
  Applied Network Science}, vol.~4, p.~4, Jan 2019.

\bibitem{barabasi1999emergence}
A.-L. Barab{\'a}si and R.~Albert, ``Emergence of scaling in random networks,''
  {\em science}, vol.~286, no.~5439, pp.~509--512, 1999.

\bibitem{fortunato2016community}
S.~Fortunato and D.~Hric, ``Community detection in networks: A user guide,''
  {\em Physics reports}, vol.~659, pp.~1--44, 2016.

\bibitem{newman}
M.~E. Newman, ``Mixing patterns in networks,'' {\em Physical Review E}, 2003.

\bibitem{peel}
L.~Peel, J.-C. Delvenne, and R.~Lambiotte, ``Multiscale mixing patterns in
  networks,'' {\em Proceedings of the National Academy of Sciences}, 2018.

\bibitem{estrada2008clumpiness}
E.~Estrada, N.~Hatano, and A.~Gutierrez, ``‘clumpiness’ mixing in complex
  networks,'' {\em Journal of Statistical Mechanics: Theory and Experiment},
  2008.

\bibitem{mcpherson2001birds}
M.~McPherson, L.~Smith-Lovin, and J.~M. Cook, ``Birds of a feather: Homophily
  in social networks,'' {\em Annual review of sociology}, 2001.

\bibitem{moody2001race}
J.~Moody, ``Race, school integration, and friendship segregation in america,''
  {\em American journal of Sociology}, vol.~107, no.~3, pp.~679--716, 2001.

\bibitem{shrum1988friendship}
W.~Shrum, N.~H. Cheek~Jr, and S.~MacD, ``Friendship in school: Gender and
  racial homophily,'' {\em Sociology of Education}, pp.~227--239, 1988.

\bibitem{feng2020mixing}
S.~Feng and A.~Kirkley, ``Mixing patterns in interdisciplinary co-authorship
  networks at multiple scales,'' {\em Scientific Reports}, vol.~10, no.~1,
  pp.~1--11, 2020.

\bibitem{rabbany2017beyond}
R.~Rabbany, D.~Eswaran, A.~W. Dubrawski, and C.~Faloutsos, ``Beyond
  assortativity: proclivity index for attributed networks (p ro n e),'' in {\em
  Pacific-Asia Conference on Knowledge Discovery and Data Mining}, Springer,
  2017.

\bibitem{pelechrinis2016va}
K.~Pelechrinis and D.~Wei, ``Va-index: Quantifying assortativity patterns in
  networks with multidimensional nodal attributes,'' {\em PloS one}, vol.~11,
  no.~1, p.~e0146188, 2016.

\bibitem{noldus2015assortativity}
R.~Noldus and P.~Van~Mieghem, ``Assortativity in complex networks,'' {\em
  Journal of Complex Networks}, vol.~3, no.~4, pp.~507--542, 2015.

\bibitem{allen2017two}
A.~Allen-Perkins, J.~M. Pastor, and E.~Estrada, ``Two-walks degree
  assortativity in graphs and networks,'' {\em Applied Mathematics and
  Computation}, vol.~311, pp.~262--271, 2017.

\bibitem{ngo2020transsortative}
S.-C. Ngo, A.~G. Percus, K.~Burghardt, and K.~Lerman, ``The transsortative
  structure of networks,'' {\em Proceedings of the Royal Society A}, vol.~476,
  no.~2237, p.~20190772, 2020.

\bibitem{lee2017homophily}
E.~Lee, F.~Karimi, C.~Wagner, H.-H. Jo, M.~Strohmaier, and M.~Galesic,
  ``Homophily and minority size explain perception biases in social networks,''
  {\em arXiv preprint arXiv:1710.08601}, 2017.

\bibitem{altenburger2018monophily}
K.~M. Altenburger and J.~Ugander, ``Monophily in social networks introduces
  similarity among friends-of-friends,'' {\em Nature human behaviour}, vol.~2,
  no.~4, pp.~284--290, 2018.

\bibitem{cantwell2019mixing}
G.~T. Cantwell and M.~Newman, ``Mixing patterns and individual differences in
  networks,'' {\em Physical Review E}, vol.~99, no.~4, p.~042306, 2019.

\bibitem{gutierrez2019multi}
L.~Guti{\'e}rrez-G{\'o}mez and J.-C. Delvenne, ``Multi-hop assortativities for
  network classification,'' {\em Journal of Complex Networks}, vol.~7, no.~4,
  pp.~603--622, 2019.

\bibitem{zachary1977information}
W.~W. Zachary, ``An information flow model for conflict and fission in small
  groups,'' {\em Journal of anthropological research}, 1977.

\bibitem{peel2017ground}
L.~Peel, D.~B. Larremore, and A.~Clauset, ``The ground truth about metadata and
  community detection in networks,'' {\em Science advances}, vol.~3, no.~5,
  p.~e1602548, 2017.

\bibitem{sapiezynski2019interaction}
P.~Sapiezynski, A.~Stopczynski, D.~D. Lassen, and S.~Lehmann, ``Interaction
  data from the copenhagen networks study,'' {\em Scientific Data}, vol.~6,
  no.~1, pp.~1--10, 2019.

\bibitem{Traud100Facebook}
A.~L. Traud, P.~J. Mucha, and M.~A. Porter, ``Social structure of facebook
  networks,'' {\em CoRR}, vol.~abs/1102.2166, 2011.

\bibitem{cinelli2019network}
M.~Cinelli, L.~Peel, A.~Iovanella, and J.-C. Delvenne, ``Network constraints on
  the mixing patterns of binary node metadata,'' {\em arXiv preprint
  arXiv:1908.04588}, 2019.

\bibitem{citraro2020identifying}
S.~Citraro and G.~Rossetti, ``Identifying and exploiting homogeneous
  communities in labeled networks,'' {\em Applied Network Science}, vol.~5,
  no.~1, pp.~1--20, 2020.

\end{thebibliography}

\subsection*{Authors}

\textbf{Giulio Rossetti} is a permanent researcher at the Information Science and Technology Institute of the Italian National Research Council (ISTI-CNR). Giulio holds a PhD in Computer Science from the University of Pisa (2015). Since 2011, he has been performing research in the fields of complex network analysis and data science as a member of the Knowledge Discovery and Data Mining laboratory. His recent work focuses on the modeling and study of dynamics of and on networks and on the definition of data-driven models for the forecast of rare events. He is the main contributor of several open-source software applications (https://github.com/GiulioRossetti), some of which have been developed to support big data analysis carried out in a series of EU projects. 
\\ \ \\
\noindent \textbf{Salvatore Citraro} is a PhD student in Computer Science at the University of Pisa and a member of the Knowledge Discovery and Data Mining Laboratory (KDD-Lab), a joint research group with the Information Science and Technology Institute of the Italian National Research Council in Pisa. Currently, his main research interests focus on complex networks, including classic and feature-rich network data analysis, with a focus on attribute-aware community discovery tasks. His PhD line of research will also include the study of linguistic networks and how their hidden structures and properties can be meaningfully exploited and applied in cognitive contexts.
\\ \ \\
\noindent \textbf{Letizia Milli} received in 2018 the PhD in Computer Science from the University of Pisa with a thesis on the study of spreading phenomena over complex networks. She is a post doc researcher in Computer Science at the University of Pisa and a member of the Knowledge Discovery and Data Mining Laboratory (KDD-Lab), a joint research group with the Information Science and Technology Institute of the Italian National Research Council in Pisa. Her research interests include data mining, quantification, diffusion of phenomena, and innovation in complex networks and the Science of Success.
\end{document}